\documentclass[12pt,preprint]{aastex}
\begin{document}
\input colordvi

\title{The diffuse GeV-TeV $\gamma$-ray emission of the Cygnus region}

\author{Xiao-Jun Bi\altaffilmark{1,2}, Tian-Lu Chen\altaffilmark{3,4}, Yue Wang\altaffilmark{1} and
        Qiang Yuan\altaffilmark{1,4}}

\affil{$^1$Key Laboratory of Particle Astrophysics, Institute of High
Energy Physics, Chinese Academy of Sciences, Beijing 100049, P. R. China\\
$^2$Center for High Energy Physics, Peking University, Beijing
100871, P.R.China\\
$^3$Physics Department of Science School, Tibet University,
Lhasa 850000, P. R. China\\
$^4$The Key Laboratory of Cosmic Rays, Ministry of Education,
Lhasa 850000, P. R. China
}

\begin{abstract}


Recently the Milagro experiment observed diffuse multi-TeV gamma
ray emission in the Cygnus region, which is significantly stronger
than what predicted by the Galactic cosmic ray model. However, the
sub-GeV observation by EGRET shows no excess to the prediction
based on the same model. This TeV excess implies possible high
energy cosmic rays populated in the region with harder spectrum
than that observed on the
Earth. 
In the work we studied this theoretical speculation in detail. We
find that, a diffuse proton source with power index
$\alpha_p\lesssim 2.3$, or a diffuse electron source with power index 
$\alpha_e\lesssim2.6$ can reproduce the Milagro's observation
without conflicting with the EGRET data. Further detections
on neutrinos, a diagnostic of  the hadronic model, and hard X-ray
synchrontron radiation, a diagnostic of the lepton model, help to
break this degeneracy. In combination with the gamma ray
observations to several hundred GeV by Fermi, we will be able to
understand the diffuse emission mechanisms in the Cygnus region
better.

\end{abstract}


\section{Introduction}

The Galactic diffuse gamma-ray emission provides important
information on the origin and propagation of the Galactic cosmic
rays (GCRs) \citep{2000ApJ...537..763S,2004ApJ...613..962S}. In
general, there exist three possible mechanisms for the diffuse
gamma-ray emission: decay of neutral pion produced by hadronic
interaction between cosmic ray (CR) nuclei and interstellar gas;
inverse Compton (IC) scattering between CR electron and
interstellar radiation field (ISRF); and the bremsstrahlung
radiation by interaction between CR electron and interstellar gas.
In addition, dark matter annihilation is emerging as an
alternative possibility to the diffuse $\gamma$-rays
\citep{2005NewAR..49..213D,2005A&A...444...51D,2007arXiv0711.1912D,
2008PhRvD..78d3001B,2008PhLB..668...87B}.

The Cygnus region, defined as
$65^{\circ}<l<85^{\circ}$ and $-3^{\circ}<b<3^{\circ}$ following
the Milagro measurements,
is rich in molecular clouds and is one of the richest star
formation regions in the Galaxy
\citep{1996ApJ...466..282D,2006A&A...458..855S}. Observations in
radio \citep{1952AuSRA...5...17P}, infrared
\citep{2000A&A...360..539K,2002A&A...389..874C,2003ApJ...597..957H},
optical \citep{1969A&A.....1..270D}, X-ray
\citep{1967ApJ...148L.119G} and $\gamma$-ray
\citep{1996A&AS..120C.423C} bands found many interesting sources
in this region. These sources have provided us valuable
information about the astrophysical processes in this region.
Ground-based very high energy (VHE) $\gamma$-ray observatories
also detected TeV $\gamma$-ray emissions in the Cygnus region
\citep{2002A&A...393L..37A, 2007ApJ...658L..33A}. Such detection
is of great importance for CR physics, since it indicates the
existence of high energy CR accelerators. Especially the Milagro
experiment\footnote{Milagro homepage,
http://umdgrb.umd.edu/cosmic/milagro.html} has made remarkable
progress recently.

The Milagro experiment discovered two sources together with a
diffuse TeV $\gamma$-ray emission in the Cygnus region 
\citep{2007ApJ...658L..33A,2008arXiv0805.0417A,2007ApJ...664L..91A}. 
For the two sources, MGRO J2031+41 is possibly the counterpart of the
unidentified TeV source TeV J2032+4130, first discovered by HEGRA
\citep{2002A&A...393L..37A}, and MGRO J2019+37 is observed for the
first time without an obvious counterpart. There are extensive
discussions about the nature of the two TeV sources
\citep{2002A&A...393L..37A,2003ApJ...597..494B,2003ApJ...589..487M,
2007PhRvD..75h3001B}. However, no consensus has been reached yet.
As for the diffuse $\gamma$-ray emission, the Milagro measurement
shows an evident excess (hereafter ``TeV excess'') compared with
the diffuse emission predicted by the Galactic cosmic ray model
GALPROP \citep[$\sim7$ times of the conventional model prediction
and $\sim2$ times of the optimized model
prediction,][]{2008arXiv0805.0417A}. On the other hand, the
sub-GeV band measurements by EGRET are well consistent with the
GALPROP prediction \citep{1997ApJ...481..205H,2007ApJ...658L..33A}
\footnote{It should be pointed out that, there are ``GeV
excesses'' of EGRET observations compared with the conventional
model of diffuse $\gamma$-ray emission in $\gtrsim$GeV energy band
\citep{1997ApJ...481..205H}. However, since it is not the main
purpose of the present work, we just briefly discuss the ``GeV
excess'' problem for the sake of completeness and do not intend to
go deep in this topic.}.

The ``TeV excess'' may be due to some unidentified TeV sources or
diffuse high energy proton or electron population  in the Cygnus
region\citep{2007ApJ...658L..33A}. If the latter is true we can
draw two general conclusions according to the Milagro and EGRET
data: 1), the spectrum of the proton or electron should be harder
than the one observed at the Earth; 2), the CRs of the source
population need to reside in the Cygnus region in order not to
exceed the locally measured CR fluxes. In this work we investigate
this possible explanation in detail by building realistic models
and deriving their implications for future experiments.

This paper is organized as follows: in Sec.2 we give an
introduction to the diffuse $\gamma$-ray emission predicted by
conventional GCR model. In Sec.3 we present a brief introduction
to the ``GeV excess'' problem. Our models to solve the ``TeV
excess'' problem and possible observational effects for future
experiments are discussed in Sec.4. Finally the conclusion is
drawn in Sec.5.

\section{Diffuse $\gamma$-rays from the Conventional GCR model}

The GCRs propagate diffusively in the Galactic magnetic field
\citep[see e.g.,][]{1990cup..book.....G}. The interactions between
CRs and the interstellar medium (ISM) can generate diffuse
$\gamma$-rays during propagation. A numerical solution of the
diffusion equation is developed by \cite{1998ApJ...509..212S},
which is known as the GALPROP model. In GALPROP the model
parameters are adjusted to reproduce the locally measured CR
spectra. The realistic distributions of ISM and ISRF are
incorporated in GALPROP to calculate the fragmentations and energy
losses. Since the realistic astrophysical inputs and the general
treatments of relevant physical processes, such as convection and
reacceleration, the GALPROP is recognized as the best CR
propagation model at present.

\begin{figure}[hbt!]
\begin{center}
\includegraphics[scale=0.6]{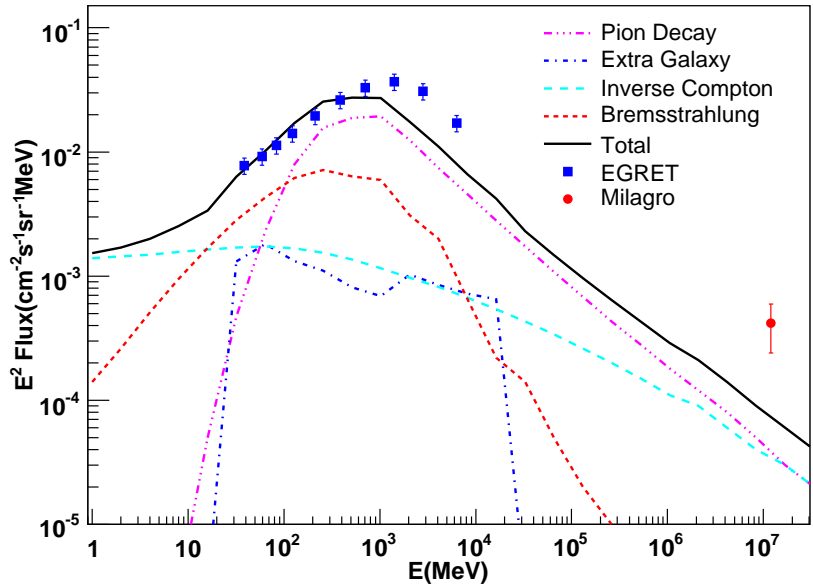}
\caption{The diffuse $\gamma$-ray spectrum in the Cygnus region predicted by
the conventional GALPROP model compared with the observational data from
EGRET \citep{2005ApJ...621..291C} and Milagro \citep{2007ApJ...658L..33A}.}
\label{fig:bk}
\end{center}
\end{figure}

We adopt the conventional GALPROP model to calculate the GCRs
propagation and the $\gamma$-ray emission (hereafter we denoted
this component as ``GCR background''). 
The isotropic extra Galactic (EG) component of diffuse
$\gamma$-rays is included
\citep{1998ApJ...494..523S,2004ApJ...613..956S}. The  results are
shown in Fig.\ref{fig:bk}. The EGRET data are adopted from the
diffuse sky maps with the point sources from 3EG catalog
subtracted \citep{2005ApJ...621..291C}. The figure shows that the
GALPROP prediction matches well with the low energy measurements
of EGRET data (except the ``GeV excess''), while for the TeV data
of Milagro, it shows a remarkable under-estimation. The model
prediction is about $6$ times lower than observation. This is
consistent with the results given in \cite{2008arXiv0805.0417A}.
It has been shown that even the optimized GALPROP model, which is
developed to solve the EGRET ``GeV excess'' problem (see the next
section), predicts only a half of the Milagro data
\citep{2007ApJ...658L..33A,2008arXiv0805.0417A}. Therefore there
must be an additional photon component with harder spectrum in the
Cygnus region. We also note that the local high density of ISM or
ISRF in the Cygnus region can not account for this ``TeV excess''
without violating the EGRET observations, since it just makes the
curves shift upwards.

\section{ The GeV excess}

The Galactic diffuse $\gamma$-rays measured by EGRET show an
excess for energies $\gtrsim$GeV compared with predictions by the
conventional GCR model, which is referred as ``GeV excess''
problem \citep{1997ApJ...481..205H}. The ``GeV excess'' appears in
all directions of the sky, and there is no special structure in
the Cygnus region. Many models are proposed to solve the ``GeV
excess'' problem by tuning the CR spectra, such as the harder
nucleon spectrum model
\citep{1997A&A...318..925G,1997ApJ...478..225M}, the harder
electron spectrum model \citep{1997JPhG...23.1765P,
1998ApJ...507..327P} or the combination of the two models
\citep{2000ApJ...537..763S}. However, these models usually cannot
reproduce the all-sky $\gamma$-ray data
\citep{2004ApJ...613..962S}.

\cite{2004ApJ...613..962S} have developed an ``optimized'' model
by adjusting the interstellar proton and electron intensities to
reproduce the $\gamma$-ray data and CR data, such as B/C,
simultaneously. The optimized model gives a good fitting to the
EGRET data of all sky directions. However, large spatial
fluctuations of proton and electron fluxes have to be incorporated
in their model \citep{2004ApJ...613..962S}.

Another approach to the ``GeV excess'' problem is by DM
annihilation \citep{2005NewAR..49..213D,
2005A&A...444...51D,2007arXiv0711.1912D}. The EGRET data in all
directions are in good agreement with predictions after taking the
DM annihilation into account if assuming a supersymmetric DM with
mass $m_{\chi}\sim50-70$ GeV \citep{2005A&A...444...51D}.
\cite{2008PhRvD..78d3001B, 2008PhLB..668...87B} extended the DM
scenario using DM subhaloes to account for the ``boost factor''
and calculate both the CR background and DM signals in more
realistic propagation models. In the present work we adopt the
model of \cite{2008PhRvD..78d3001B} to fit the GeV data. However,
it should be noted that it is far away from the final answer to
the ``GeV excess'' problem. To go details of this issue is beyond
the scope of the present study.

\section{TeV excess}

In this section we turn to the ``TeV excess'' problem. We
introduce a high energy population of CRs to explain the Milagro
data. It is known that the Cygnus region is rich in potential CR
accelerators, such as the Wolf-Rayet stars
\citep{2001NewAR..45..135V}, OB associations
\citep{1985Ap&SS.108..237B} and supernova remnants
\citep[SNRs,][]{2004BASI...32..335G}. Here we assume there exist a
(or several) CR accelerating source(s) in this region. The ages of
the sources should not be too old (e.g., $10^6\sim 10^7$ yr, for
typical diffuse coefficient $D\sim10^{28}$ cm$^2$ s$^{-1}$, the
propagation length after this time is about several hundred pc) so
that the CRs can distribute diffusively in the Cygnus region and
will not affect the measurements of CR fluxes at the Earth. We
discuss in detail the hadronic and leptonic scenarios in the
following.

\subsection{The hadronic model}

We first consider a proton source uniformly distributed in the
Cygnus region. We assume it has the energy spectrum as ${\rm
d}\phi_p/{\rm d}E_p\propto E_p^{-\alpha_p}\exp(-E_p/E_p^c)$. The
spectrum index $\alpha_p$ is adopted as $2.3$, which can well
reproduce the Milagro data and is still consistent with the low
energy data by EGRET. It should be noted that the adoption of
$\alpha_p$ is not arbitrary. On one hand, it cannot be too soft in
order not to exceed the sub-GeV observations by EGRET. On the
other hand, the Fermi acceleration of CRs from shock waves
predicts spectrum index $\sim 2$ \citep{1990cup..book.....G}; and
the diffusion in the interstellar medium can soften the spectrum a
bit (e.g., $1/3$ for a Kolmogorov diffusion). Therefore the
Milagro observations can indeed limit the source spectrum in a
relative narrow range. The cut-off energy $E_p^c$ is due to the
acceleration limit of the source. For supernova blast shock it is
known to be $100\sim 1000$ TeV for protons
\citep{1996APh.....5..367B}. The shock acceleration by stellar
winds of red giants and Wolf-Rayet stars gives $10$ PeV
\citep{1988ApJ...333L..65V}. The ensemble shocks of OB association
can even give the maximum energy about $100$ PeV
\citep{1990ZhETF..98.1255B}. Here we adopt the cut-off energy to
be $1\sim 10$ PeV. Such a cut-off seems not to affect the results
of $\gamma$-rays at $\sim10$ TeV. However, it will affect the high
energy neutrino flux, as shown below. The normalization of the CR
proton density is tuned to be consistent with the high energy
observations by Milagro.

\begin{figure}[hbt!]
\begin{center}
\includegraphics[scale=0.6]{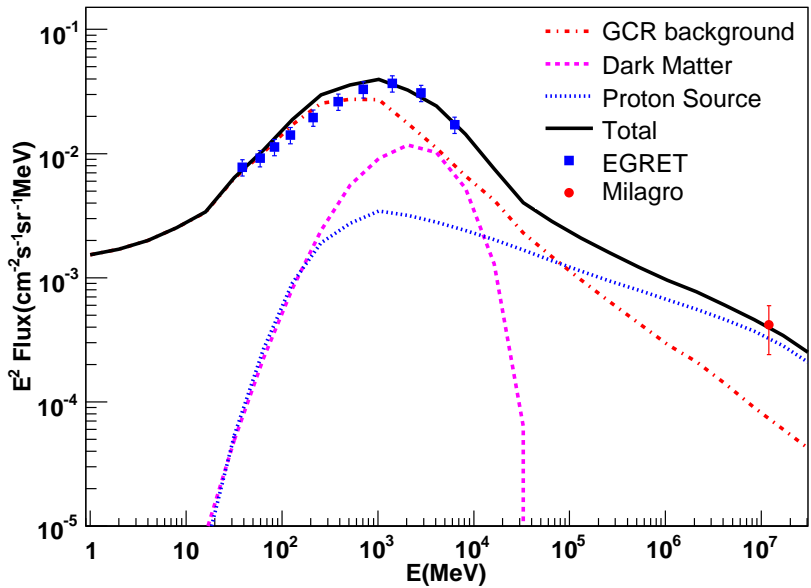}
\caption{The energy spectrum of diffuse $\gamma$-rays in the Cygnus
region as measured by EGRET and Milagro and predicted by the GALPROP
model combined with additional proton source and DM annihilation
components. See text for details.
}\label{fig:hadron}
\end{center}
\end{figure}

The $\gamma$-ray spectrum after taking such new sources into
account is shown in Fig.\ref{fig:hadron}. It shows a good
agreement between the model prediction and observations. We notice
that the high energy behavior of $\gamma$-rays is dominated by the
new proton source contribution, so precise measurements of the
high energy spectrum (from several GeV to hundreds of TeV) will be
helpful to determine the spectrum of the source term. The total
energy of the proton source is estimated as $W(>1{\rm\
GeV})\sim10^{51}(\frac{1{\rm\ cm}^{-3}} {n_H})(\frac{d}{1{\rm\
kpc}})^2$ erg. For typical ISM density $n_H\approx 1{\rm\
cm}^{-3}$ and the distance of the source $d\sim1$ kpc \citep[which
is comparable with the molecular clouds and OB associations in
this region,] []{2006A&A...458..855S}, the total energy is similar
with the energy release of a typical core-collapse supernova
\citep{1962PASP...74...70Z}.

For inelastic $p-p$ collision, the pion mesons $\pi^+$, $\pi^-$
and $\pi^0$ are generated with almost equal amount
\citep{1990cup..book.....G}. $\gamma$-rays are produced through
the decay of $\pi^0$ mesons. The decay of $\pi^{\pm}$ will lead to
neutrino emission. In general the initial neutrino flavor ratio is
$\nu_e:\nu_{\mu}:\nu_{\tau}=1:2:0$. Because of neutrino
oscillation the flavor ratio on the Earth becomes
$\nu_e:\nu_{\mu}: \nu_{\tau}=1:1:1$. The ratio between the total
number of neutrinos and $\gamma$-ray photons is $3:1$. The typical
energy of neutrinos is $\sim1/2$ of the $\gamma$ photons, so we
have \citep{2007PhRvD..75h3001B}
\begin{equation}
\frac{{\rm d}\phi_{\nu}}{{\rm d}E_{\nu}}=2
\frac{{\rm d}\phi_{\gamma}}{{\rm d}E_{\gamma}},
\end{equation}
for each kind of neutrinos.

\begin{figure}[hbt!]
\begin{center}
\includegraphics[scale=0.6]{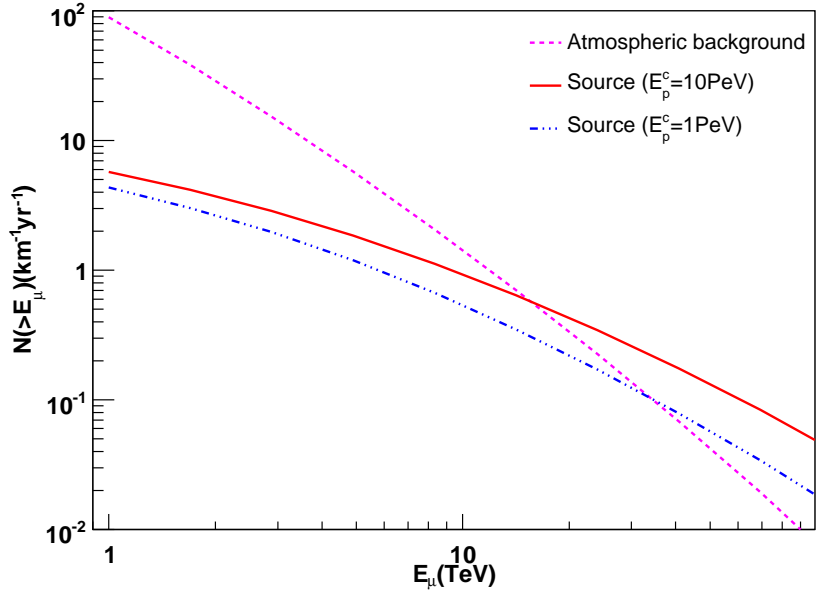}
\caption{The cumulative muon events induced by neutrinos from the Cygnus
region and atmospheric background on IceCube for one-year observation.}
\label{fig:neutrino}
\end{center}
\end{figure}

In Fig. \ref{fig:neutrino} we show the cumulative muon flux
induced by $\nu_{\mu}+\bar{\nu}_{\mu}$ on a km$^3$ detector, like
IceCube, for one-year observation. The muon flux including both
the containing events (that muons generated in the detector) and
the through-going events (that muons generated outside the
detector and propagates into the detector volume) is calculated
following the method of \cite{2006PhRvD..74f3007K}. The absorption
of neutrinos by the Earth along the Cygnus direction (R.A.
$\sim20^h$, Dec. $\sim40^{\circ}$) is taken into account. This
absorption effect is known to be significant for neutrino energies
greater than $\sim10$TeV \citep{2006NIMPA.567..405L}. The
background induced by the atmospheric neutrinos is also shown in
the figure, which is from a 102 deg$^2$ sky region with two
$3^{\circ}\times 3^{\circ}$ areas around the two TeV sources, TeV
J2021+4130 and MGRO J2019+37, excluded
\citep{2007ApJ...658L..33A}. The atmospheric neutrino flux is
adopted from \cite{2007PhRvD..75d3006H}, with the zenith angle
$\cos \theta_z\sim 0.6-0.7$  \footnote{The zenith angle of the
Cygnus region relative to the detector at the South pole is about
$130^{\circ}$, i.e., the neutrinos are up-going. Since there is a
mirror symmetry between the up-going and down-going atmospheric
neutrinos, it is equivalent to the atmospheric neutrino at
direction with zenith angle $\sim50^{\circ}$.}.

It can be seen from Fig. \ref{fig:neutrino} that if the cut-off
energy of protons is high enough, the neutrino induced muons are
marginally observable over the atmospheric background for energy
$>10$ TeV. For a 15-year observation of IceCube, we estimate the
significance $\sim3\sigma$ for an energy threshold $10$ TeV.
Compared with the neutrino emission from MGRO J2019+37
\citep{2007PhRvD..75h3001B}, our result of diffuse neutrinos is
$\sim 4$ times larger. This is reasonable as the total diffuse TeV
photon emission is about $4$ times larger than that from MGRO
J2019+37 \citep{2007ApJ...658L..33A}. However, since the diffuse
atmospheric background is about $30$ times higher than the case of
MGRO J2019+37, the detectability becomes worse
\citep{2007PhRvD..75h3001B}. \cite{2006PhRvD..74f3007K} got the
similar conclusion about the diffuse neutrinos.
If the cut-off energy of protons is as low as $\sim 1$ PeV, the
signal is dominated by the atmospheric background and is almost
invisible.

It should be pointed out that the neutrinos from the GCRs
cosmic rays interacting with the ISM are negligible and not shown
here. From Fig.\ref{fig:bk} we can see that the $\pi^0$
component of $\gamma$-ray flux from GCR background is about one
order of magnitude lower than the Milagro data. Therefore the
neutrino emission from this component is also very small. It can
also be noted that the diffuse $\gamma$-ray emission is actually
not uniformly distributed in the Cygnus region, but concentrates
in some areas. Therefore the atmospheric background will be lower
if proper observational region is concentrated, and then the
detectability of signals will be improved.

\subsection{The leptonic model}

A high energy electron population similar with the hadronic one is also
able to produce the TeV $\gamma$-rays through IC scattering
\citep{2006A&A...449..223A}. Similar with the hadronic model, an electron
source population with differential energy spectrum ${\rm d}\phi_e/{\rm d}
E_e\propto E_e^{-\alpha_e}\exp(-E_e/E_e^c)$ is introduced to account for
the TeV $\gamma$-ray emission. The high energy spectral index of the
electron population is adopted as $\alpha_e=2.6$, which is harder than
the Galactic background electron spectrum $\sim3.3$. The cut-off energy
is adopted as $E_e^c\sim 100-1000$ TeV. Note that due to the fast 
energy losses of electrons in interstellar magnetic field, the maximum 
energy of electrons generally cannot be higher than several hundred TeV
\citep{1990cup..book.....G}. 

However, such an electron population
will contribute too many low energy $\gamma$-rays through
bremsstrahlung radiation and will exceed the EGRET data. Therefore
a broken power-law of spectral index is introduced to match the
 data. We adopt $\alpha_e=1.5$ for energy lower than
$4$ GeV, which is similar to that done in GALPROP
\citep{2004ApJ...613..956S}.

For the high energy slope, it is not arbitrary due to observations
and theoretical arguments. It is known that the fast energy loss
through synchrotron and IC will soften the power index of electron
spectrum by about $1$. Therefore if the source spectrum is $\sim
2$ as expected from 1st order Fermi acceleration, the propagated
one is $\sim 3$. However, we argue that the reacceleration of
electrons by the stellar wind shock of massive stars in the Cygnus
region can maintain a harder spectrum
--- $2.6$ as we adopted.

\begin{figure}[hbt!]
\begin{center}
\includegraphics[scale=0.6]{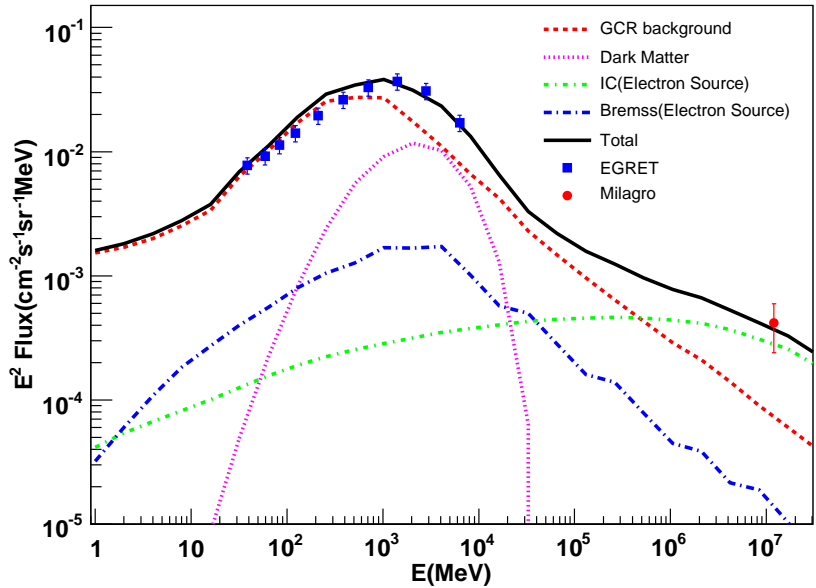}
\caption{The energy spectrum of diffuse $\gamma$-rays in the Cygnus
region as measured by EGRET and Milagro and predicted by the GALPROP
model combined with additional electron source and DM annihilation
components. See text for details.}
\label{fig:lepton}
\end{center}
\end{figure}

The target ISRF field is adopted as the combination of three
components: CMB, infrared from dust and optical from stars
\citep{2005ICRC....4...77P}. The predicted $\gamma$-ray spectrum
is shown in Fig.\ref{fig:lepton}. It shows that this model can
also give good description to data. The total energy of this
electron source is estimated as $W(>1{\rm\ GeV})\sim
10^{49}(\frac{d}{1{\rm\ kpc}})^2$ erg.

\begin{figure}[hbt!]
\begin{center}
\includegraphics[scale=0.6]{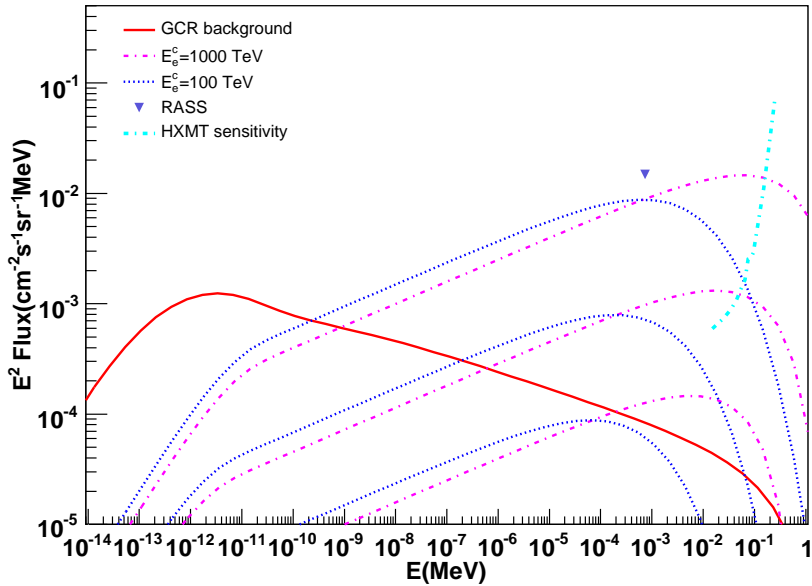}
\caption{The synchrotron radiation from the electron source population
for magnetic fields $B=1,\,3$ and $10\mu$G (from bottom to top)
respectively. The GCR background synchrotron radiation is calculated
using GALPROP. The observed X-ray limit is from RASS
\citep{1997ApJ...485..125S}. Also shown is the sensitivity
curve for fix-direction observation of HXMT.}
\label{fig:syn}
\end{center}
\end{figure}

It is expected that such a high energy electron population will
generate low energy synchrotron radiation in the interstellar
magnetic field. We show the synchrotron radiation in
Fig.\ref{fig:syn} for the magnetic field $B=1,\,3$ and $10\ \mu$G
respectively. The synchrotron spectrum is
$\sim(\alpha_e+1)/2=1.8$, which is significantly different from
the one by GCR background electrons, as shown in
Fig.\ref{fig:syn}.

Compared with the GCR background contribution, the additional 
electrons will contribute high energy synchrotron
radiation, and might be detected by X-ray detectors. We find in
the soft X-ray background (SXRB) from ROSAT All-Sky Survey
\citep[RASS,][]{1997ApJ...485..125S} that there is no distinct
excess in the Cygnus region compared with nearby sky regions. The
result from RASS in the Cygnus region is shown in this figure as
an upper limit of the synchrotron radiation. It is shown that even
for the magnetic field $10\ \mu$G the predicted synchrotron
radiation in the soft X-ray band is consistent with the
observational SXRB. However, the peak of the synchrotron radiation
is in the hard X-ray band ($\sim$ hundred keV) if the electron
source has an energy cut $\sim 1$ PeV. It may be detectable by the
future satellite Hard X-ray Modulation Telescope \citep[HXMT;][]
{2002cosp...34E..16W,2006cosp...36.2815L}. HXMT is a space
satellite with high sensitivity and a field of view
$5.7^{\circ}\times5.7^{\circ}$, which is devoted to performing a
hard X-ray all-sky imaging survey. The wide field of view and high
sensitivity make it possible to detect the diffuse X-ray emission
from the Cygnus region. The sensitivity curve of HXMT for the
fix-direction observation of diffuse emission is shown in
Fig.\ref{fig:syn}. If the astrophysical background in hard X-ray
band is not too high, the diffuse emission in the Cygnus region is
detectable. If $E_e^c=100$ TeV, the ``TeV excess'' of Milagro data
can also be reproduced, but the synchrotron radiation is almost
invisible due to the high SXRB.

\section{Conclusions and Discussions}

The TeV $\gamma$-ray emission of the Cygnus region observed by
Milagro shows significant excess compared with the conventional CR
model. In this work we introduce a high energy proton or electron
source to explain this high energy $\gamma$-ray emission. It is
shown that both the hadronic and leptonic models can give good
explanation to the current data. The total energy needed for the
source population is consistent with the typical energy output of
a supernova.

The associated neutrino emission for the hadronic model and
synchrotron radiation for the leptonic model are discussed as
possible probes to discriminate these two scenarios. For the
hadronic model,
we estimate a $3\sigma$ significance for 15-year exposure of
IceCube in case the protons cutoff energy reaches $\sim10$ PeV.
The non-null result of the neutrino detection will support the
hadronic model. For the leptonic scenario, the synchrotron
radiation by the electrons is more luminous at the X-ray
wavelength.
We show that if the energy cut-off of the electron source is as
high as $\sim1$ PeV, the synchrotron radiation will peak in the
hard X-ray bands and it might be observable on the satellite HXMT.

In the calculation, the ISM density, the ISRF intensity and the
magnetic field are adopted as typical Galactic values. Since there
are a large number of massive stars and molecular clouds in the
Cygnus region, the ISRF and ISM density may be higher than the
Galactic average values. We should point out that this does not
change the basic conclusion in this work. This is because a higher
ISM density or ISRF intensity can be effectively compensated by a
lower CR flux. For the hadronic model, the neutrino flux will keep
the same since neutrino flux is proportional to the $\gamma$-ray
flux. For the leptonic model a higher ISRF intensity will correspond 
to a lower electron luminosity. However,
considering the magnetic field in the Cygnus region may be also
stronger than the Galactic value we may expect similar synchrotron
radiation as given in the work.

Finally, since the present data are lack in the energy interval
from $\sim10$GeV to $\sim1$TeV due to the transition of detection
technology from space to ground there are large uncertainties for
model construction. The space telescope Fermi will extend the
detection energy up to $\sim300$GeV \citep{2004HEAD....8.2101R}.
Together with some ground-based experiments, such as ARGO, which
may lower the energy threshold down to $\sim100$GeV
\citep{1999NuPhS..78...38B}, we will have better understanding in
the emission mechanism of this region in the future.

\acknowledgments

We thank Lu Ye, Lu Fang-Jun, Qu Jin-Lu and Zhang Shuang-Nan for helpful
discussions. The work is supported by the NSF of China under the grant
Nos. 10575111, 10773011 and supported in part by the Chinese Academy of
Sciences under the grant No. KJCX3-SYW-N2.


\bibliographystyle{aa}
\bibliography{gammaray}

\end{document}